\documentclass[twocolumn,showpacs,showkeys]{revtex4}
\usepackage{graphicx}
\input{psfig.sty}

\newcommand{\bastar}{\begin{eqnarray*}}
\newcommand{\eastar}{\end{eqnarray*}}
\newskip\humongous \humongous=0pt plus 1000pt minus 1000pt

\newif\ifdtup

\relax
\newcommand{\be}{\begin{equation}}
\newcommand{\ee}{\end{equation}}
\newcommand{\bea}{\begin{eqnarray}}
\newcommand{\eea}{\end{eqnarray}}

\newcommand{\n}{\hat n}

\newcommand{\dfrac}{\displaystyle\frac}
\newcommand{\ba}{\begin{array}}
\newcommand{\ea}{\end{array}}

\newcommand{\nn}{\nonumber}
\newcommand{\hn}{\hat n}
\begin{document}
\title{Non-Abrikosov Vortex and Topological Knot in Two-gap Superconductor}
\bigskip
\author{Y. M. Cho}
\email{ymcho@yongmin.snu.ac.kr}
\author{Pengming Zhang}
\email{zhpm@phya.snu.ac.kr}
\affiliation{Center for Theoretical Physics and School of Physics \\
College of Natural Sciences, Seoul National University,
Seoul 151-742, Korea  \\
}
\begin{abstract}
~~~We establish the existence of topologically stable
knot in two-gap superconductor whose topology $\pi_3(S^2)$
is fixed by the Chern-Simon index of the electromagnetic potential.
We present a helical magnetic vortex
solution in Ginzburg-Landau theory of two-gap superconductor
which has a non-vanishing condensate at the core,
and identify the knot as a twisted magnetic
vortex ring made of the helical vortex.
We discuss how the knot can be constructed in
the recent two-gap $\rm MgB_2$ superconductor.
\end{abstract}
\pacs{74.20.-z, 74.20.De, 74.60.Ge, 74.60.Jg, 74.90.+n}
\keywords{twisted magnetic vortex ring, topological knot in
two-gap superconductor, Chern-Simon topology of magnetic field}
\maketitle

Topological objects, in particular finite energy topological
objects have played important role
in physics. In condensed matter
physics the best known topological object is the Abrikosov
vortex in one-gap superconductors (and similar ones in Bose-Einstein
condensates and superfluids), which have been the subject
of intensive studies.   
But the advent of two-component BEC 
and two-gap superconductors \cite{bec,sc}
has opened a new possibility for us to have far more
interesting topological objects
in condensed matters \cite{ijpap,prb05,ruo,pra05,baba}.
{\it The purpose of this Letter is two-fold;
to present a new type of non-Abrikosov
vortex which has a non-vanishing condensate at the core
and to establish the existence of a stable knot
in Abelian two-gap superconductors.}
We first construct a new type of non-Abrikosov
vortex which has a non-vanishing condensate at the core
in Ginzburg-Landau theory of two-gap superconductor,
and show that it can be twisted to form a helical magnetic vortex. 
With this we show that the twisted vortex ring
made of the helical magnetic vortex becomes a topological knot
whose knot topology $\pi_3(S^2)$ is fixed by
the Chern-Simon index of the electromagnetic potential.
The knot has two supercurrents, the electromagnetic current
around the knot tube which confines the magnetic flux
and the neutral current along the knot which
generates a net angular momentum which prevents 
the collapse of the knot,
This means that the knot has a dynamical (as well as
topological) stability.

There have been two objections against the existence of 
a stable magnetic vortex ring in Abelian superconductors. 
First, it is supposed to be unstable
due to the tension created by the ring \cite{huang}.
Indeed if one contructs a vortex ring from an Abrikisov vortex,
it becomes unstable because of the tension.
On the other hand, if we first twist the magnetic vortex
making it periodic in $z$-coordinate and then
make a vortex ring connecting the periodic ends together,
it should become a stable vortex ring. This is
because the non-trivial twist of the magnetic field
forbids the untwisting of the vortex ring
by any smooth deformation of field configuration.
The other objection is that the Abelian gauge theory
is supposed to have no non-trivial knot topology
which alows a stable vortex ring.
This again is a common misconception. Actually
the theory has a well-defined knot
topology $\pi_3(S^2)$ described by the Chern-Simon index
of the electromagnetic potential. 
So {\it a priori} there is
no reason whatsoever why the Abelian superconductor can not have
a topological knot.
The real question then is whether the dynamics
of the superconductor can actually allow
such a knot configuration.
In the following we show that indeed a two-gap
Abelian superconductor does.

In mean field approximation
the Hamiltonian of the Ginzburg-Landau theory of two-gap
superconductor could be expressed by \cite{baba}
\bea
&{\cal H} = \dfrac{\hbar^2}{2m_1} |(\mbox{\boldmath $\nabla$}
+ ig \mbox{\boldmath $A$}) \tilde \phi_1|^2
+\dfrac{\hbar^2}{2m_2} |(\mbox{\boldmath $\nabla$}
+ ig \mbox{\boldmath $A$}) \tilde \phi_2|^2 \nn\\
&+ \tilde V(\tilde \phi_1,\tilde \phi_2) 
+ \dfrac{1}{2} (\mbox{\boldmath $\nabla$}
\times \mbox{\boldmath $A$})^2, \label{scfe1}
\eea
where $\tilde V$ is the effective potential.
One can simplify the above Hamiltonian with a proper normalization
of $\tilde \phi_1$ and $\tilde \phi_2$ to 
$\phi_1=(\hbar/\sqrt{2m_1}/) \tilde \phi_1$ and
$(\hbar/\sqrt{2m_2}) \tilde \phi_2$,
\bea
&{\cal H} = |(\mbox{\boldmath $\nabla$}
+ ig \mbox{\boldmath $A$}) \phi|^2 + V(\phi)
+ \dfrac{1}{2} (\mbox{\boldmath $\nabla$}
\times \mbox{\boldmath $A$})^2,
\label{scfe2}
\eea
where $\phi=(\phi_1,\phi_2)$ and $V$ is 
the normalized potential. A most general quartic potential
which has the $U(1)\times U(1)$ symmetry is given by
\bea
&V =\dfrac{\lambda_{11}}{2}|\phi_1|^4+\lambda_{12}|\phi_1|^2
|\phi_2|^2+\dfrac{\lambda_{22}}{2}|\phi_2|^4 \nn\\
&-\mu_1|\phi_1|^2-\mu_2|\phi_2|^2,
\label{scpot1}
\eea
where ${\lambda}_{ij}$ are the quartic coupling
constants and $\mu_i$ are the chemical potentials.
But in this paper we will adopt the following potential
for simplicity
\bea
&V=\dfrac{\lambda}{2} \Big((\phi^{\dagger} \phi)
-\dfrac{\mu^2}{\lambda} \Big)^2 
+ \alpha |\phi_1|^2 |\phi_2|^2.
\label{scpot}
\eea
Notice that when $\alpha=0$, the Hamiltonian (\ref{scfe2}) has 
a global $SU(2)$ symmetry as well as the the local $U(1)$ 
symmetry, so that the $\alpha$-term can be viewed as
a symmetry breaking term of $SU(2)$ to $U(1)$.
We choose (\ref{scpot}) simply because the existence of 
the topological objects we discuss in this paper is not so sensitive 
to the detailed form of the potential. 

With the above Hamiltonian one may try to obtain a non-Abelian 
vortex. With 
\bea 
&\phi =\dfrac{1}{\sqrt 2} \rho
\xi,~~~~~{\xi}^{\dagger}\xi = 1, 
~~~~~\hat n = \xi^{\dagger} \vec \sigma \xi, \nn\\
&C_\mu = \dfrac{2i}{g} \xi^{\dagger}\partial _\mu \xi,
~~~~~\rho_0=\dfrac{2\mu^2}{\lambda},
\eea 
we have the following equations of motion \cite{ijpap} 
\bea
&\partial ^2 \rho - \Big( \dfrac{1}{4} (\partial _\mu \hat n)^2+
g^2
(A_\mu - \dfrac{1}{2} C_\mu)^2 \Big) \rho \nn\\
&= \dfrac{\lambda}{2} (\rho^2-\rho_0^2)\rho 
+\alpha \rho^3 \xi_1^{*}\xi_1 \xi_2^{*}\xi_2 , \nn\\
&\hat n \times \partial ^2 \hat n + 2 \dfrac{\partial_\mu \rho}{
\rho} \hat n \times \partial_\mu \hat n
- \dfrac{2}{g\rho^2} \partial_\mu F_{\mu\nu} \partial_\nu \hat n \nn\\
&=\dfrac{\alpha}{2} (\hat k \cdot \hn) 
\rho^2 \hat{k} \times \hat{n}, \nn\\
&\partial_\mu F_{\mu\nu} =g^2 \rho^2 (A_\mu - \dfrac{1}{2} C_\mu),
\label{sceq2} 
\eea 
where $\vec \sigma$ is the Pauli matrix and $\hat k=(0,0,1)$. 
Notice that it has the vacuum at 
$\rho=\rho_0$, and $\xi=(1,0)$ or $\xi=(0,1)$.

The equation allows two conserved currents,
the electromagnetic current $j_\mu$ and the
neutral current $k_\mu$, 
\bea 
&j_\mu = g^2 \rho^2(A_\mu -\dfrac 12C_\mu) \nn\\
&k_\mu = g^2 \rho^2 \Big[ A_\mu \big(\xi_1^{*}\xi_1
-\xi_2^{*}\xi_2 \big) 
+\dfrac{i}{g} \big(\partial_\mu \xi_1^{*}\xi_1 \nn\\
&+\xi_2^{*}\partial_\mu \xi_2 \big) \Big], 
\label{jkc} 
\eea 
which are nothing but the Noether
currents of the $U(1)\times U(1)$ symmetry of the Hamiltonian
(\ref{scfe2}). Indeed they are
the sum and difference of two electromagnetic currents of
$\phi_1$ and $\phi_2$
\bea
j_\mu = j^{(1)}_\mu + j^{(2)}_\mu,
~~~~~k_\mu = j^{(1)}_\mu - j^{(2)}_\mu.
\eea
Clearly the conservation of $j_\mu$ follows from 
the last equation of (\ref{sceq2}). But the conservation of 
$k_\mu$ comes from the second equation of (\ref{sceq2}),
which (together with the last equation) tells the
existence of a partially conserved $SU(2)$ current
$\vec j_\mu$,  
\bea 
\vec j_\mu = g\rho^2 \Big(\frac 12\vec{n} \times \partial_\mu
\vec{n} -g(A_\mu -\frac 12C_\mu )\vec{n} \Big).
\eea 
This $SU(2)$ current is exactly conserved when 
$\alpha=0$. But notice that $k_\mu=\hat k \cdot \vec j_\mu$.
This assures the conservation of $k_\mu$ even when $\alpha \neq 0$. 
It is interesting to notice
that $j_\mu$ and $k_\mu$ are precisely the $\hn$ and $\hat k$
components of $\vec j_\mu$.

To obtain the desired knot we construct a helical magnetic vortex
first \cite{ijpap}. Let $(\varrho,\varphi,z)$ be 
the cylindrical coodinates and let 
\bea 
&\rho=\rho(\varrho),~~~~~\xi = \Bigg( \matrix{\cos
\dfrac{f(\varrho)}{2} \exp (-in\varphi) \cr
\sin \dfrac{f(\varrho)}{2} \exp (imkz)} \Bigg), \nn\\
&A_\mu= \dfrac{1}{g} \Big (n A_1(\varrho) \partial_\mu\varphi  +
mk A_2(\varrho) \partial_\mu z \Big),
\label{scans}  
\eea  
and find  
\bea &\n=\Bigg(\matrix{\sin{f(\varrho)}\cos{(n\varphi+mkz)} \cr
\sin{f(\varrho)}\sin{(n\varphi+mkz)} \cr
\cos{f(\varrho)}}\Bigg), \nn\\
&C_\mu=\dfrac{\cos{f(\varrho)}+1}{g} n\partial_\mu \varphi +
\dfrac{\cos{f(\varrho)}-1}{g} mk\partial_\mu z, \nn\\
&j_\mu = g\rho^2 \Big(n \big(A_1-\dfrac{\cos{f}+1}{2}\big)
\partial_\mu \varphi \nn\\
&+ mk\big(A_2-\dfrac{\cos{f}-1}{2}\big)
\partial_\mu z \Big), \nn\\
&k_\mu = g\rho^2 \Big(n \big(A_1 \cos{f}-\dfrac{\cos{f}+1}{2}\big)
\partial_\mu \varphi \nn\\
&+ mk\big(A_2 \cos{f}+ \dfrac{\cos{f}-1}{2}\big)
\partial_\mu z \Big).
\label{scans1}
\eea   
With this (\ref{sceq2}) becomes  
\bea  
&\ddot{\rho}+\dfrac{1}{\varrho}\dot\rho
- \Big[\dfrac{1}{4} \Big(\dot{f}^2
+\big(\dfrac{n^2}{\varrho^2} + m^2 k^2 \big) \sin^2{f}\Big) \nn\\
&+\dfrac{n^2}{\varrho^2} \Big(A_1-\dfrac{\cos{f}+1}{2}\Big)^2
+ m^2 k^2 \Big(A_2-\dfrac{\cos{f}+1}{2}\Big)^2 \Big]\rho \nn\\
&= \dfrac{\lambda}{2}(\rho^2-\rho_0^2)\rho 
+\dfrac{\alpha}{4} \rho^3 \sin ^2 f, \nn\\
&\ddot{f} + \big(\dfrac{1}{\varrho} +2\dfrac{\dot{\rho}}{\rho}
\big)\dot{f}
-2 \Big(\dfrac{n^2}{\varrho^2}\big(A_1-\dfrac{1}{2}\big) \nn\\
&+ m^2 k^2 \big(A_2+\dfrac{1}{2}\big) \Big)\sin{f} 
= \dfrac{\alpha}{4} \rho^2 \sin 2f, \nn\\
&\ddot{A}_1-\dfrac{1}{\varrho}\dot{A}_1 -g^2 \rho^2
\Big(A_1-\dfrac{\cos{f}+1}{2}\Big) = 0, \nn\\
&\ddot{A}_2+\dfrac{1}{\varrho}\dot{A}_2 -g^2 \rho^2
\Big(A_2-\dfrac{\cos{f}-1}{2}\Big) = 0.
\label{sceq3}
\eea   
Now, we impose the following boundary condition for
the non-Abelian vortex,   
\bea
&\dot \rho(0) = 0,~~~\rho(\infty) = \rho_0, ~~~f (0) = \pi,
~~~f (\infty) = 0, \nn\\
& A_1 (0) = 0,~~~A_1 (\infty) =1, \nn\\
&\dot A_2 (0) = 0,~~~A_2 (\infty) =0.
\label{scbc}
\eea   
This need some explanation. First, notice that here we have
$\dot \rho(0) = 0$, not $\rho(0) = 0$. This would have been
unacceptable in one-gap superconductor because this would create
a singularity at the core of the vortex. But with the ansatz
(\ref{scans}) the boundary condition allows a smooth vortex in two-gap
superconductor. Secondly, we have $\dot A_2(0)=0$ in stead of
a condition on $A_2(0)$. It turns out that this is 
the correct way to obtain a smooth solution. Indeed
we find that $A_2(0)$ is determined by $k$.

\begin{figure}[t]
\includegraphics[scale=0.6]{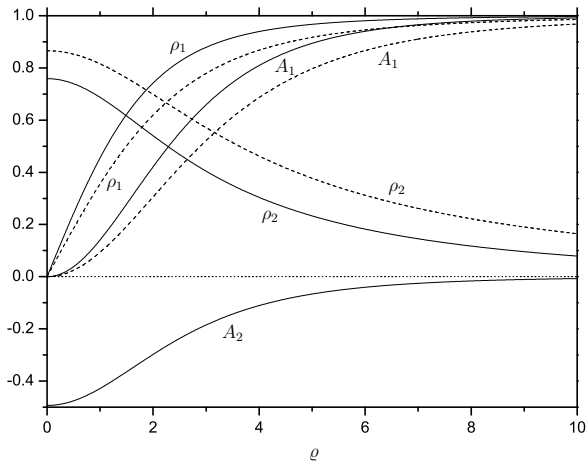}
\caption{The helical vortex (solid line) with $m=n=1$ and the
straight vortex (dotted line) $m=0$ and $n=1$ in two-gap
superconductor. Here we have put $g=1,~\lambda=2$, $k=0.1~\rho_0$, 
$~\alpha=0.01$, and $\varrho$ is in the unit of $1/\rho_0$. Notice that 
the straight vortex has no $A_2$.} 
\label{scrhoa}
\end{figure}

\begin{figure}[t]
\includegraphics[scale=0.6]{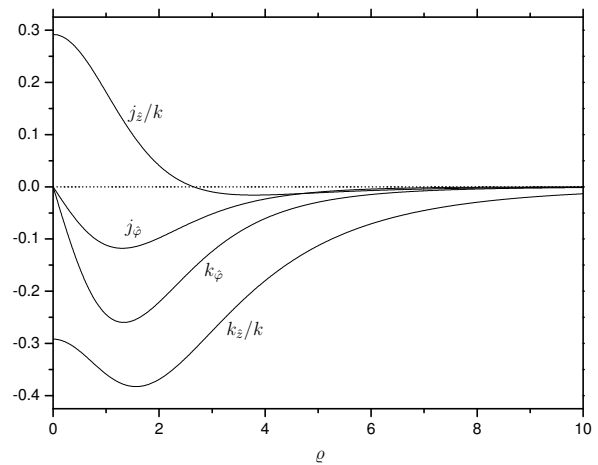}
\caption{The helical supercurrents with $m=n=1$ in 
two-gap superconductor. Here we have put
$g=1,~\lambda=2$, $k=0.1~\rho_0$, $~\alpha=0.01$, and $\varrho$ is
in the unit of $1/\rho_0$.} 
\label{2scca}
\end{figure}

With the boundary condition we can integrate (\ref{sceq3}) and
obtain the non-Abelian vortex solution of the two-gap
superconductor which is shown in Fig.\ref{scrhoa}. Notice that we
have plotted $\rho_1= \rho \cos f/2$ and $\rho_2= \rho \sin f/2$
which represent the density of $\phi_1$ and $\phi_2$, in stead of
$\rho$ and $f$, in the figure. There are three points that should
be emphasized here. First, the doublet $\phi$ starts from the
second component at the core, but the first component takes over
completely at the infinity. This is due to the boundary condition
$f(0)=\pi$ and $f(\infty)=0$, which assures that our solution
describes a new type of non-Abrikosov vortex 
which has a non-vanishing condensate at the core \cite{ijpap}.
Secondly the electromagnetic current $j_\mu$ of the vortex 
is helical, so that it creates a helical magnetic flux which
has a quantized flux $\Phi_{\hat z}=2\pi n/g$
around the $z$-axis as well as a (fractionally) quantized flux
$\Phi_{\hat \varphi}=-2\pi m A_2(0)/g$ along the $z$-axis.
And obviously the two magnetic fluxes are linked
together. But it creates a net current $i_{\hat \varphi}$ 
only around the $z$-axis, with 
$i_{\hat z}=i^{(1)}_{\hat z}+i^{(2)}_{\hat z}=0$.
Finally the vortex has another (neutral) helical
$k_\mu$, which has a net current $k_{\hat z}$ along the $z$-axis
(as well as $k_{\hat \varphi}$ around the vortex).
Notice that $k_{\hat z}=i^{(1)}_{\hat z}-i^{(2)}_{\hat z}
=2i^{(1)}_{\hat z}$. This is because 
the two electromagnetic currents
$i^{(1)}_{\hat z}$ and $i^{(2)}_{\hat z}$
of $\phi_1$ and $\phi_2$ are equal but flow oppositely, 
which generates a nonvanishing
$k_{\hat z}$. This is why we call $k_\mu$ a neutral current.
In this sense the vortex
can be viewed as superconducting, even though the net
electromagnetic current along the knot is zero.
The two helical supercurrents $j_\mu$ and $k_\mu$
are plotted in Fig. \ref{2scca}.

Of course, the helical magnetic vortex has only a heuristic value,
because it will be untwisted
unless the periodicity condition is enforced by hand.
But if we make a vortex ring by smoothly bending and connecting
two periodic ends, the periodicity condition is
naturally enforced. And the twisted magnetic vortex ring
becomes a topological knot \cite{ijpap}.
There are two ways to describe the knot topology.
First, the doublet $\xi$ (in the vortex ring)
acquires a non-trivial topology $\pi_3(S^2)$ which provides
the knot quantum number,
\bea
&Q = - \dfrac {1}{4\pi^2} \int \epsilon_{ijk} \xi^{\dagger}
\partial_i \xi ( \partial_j \xi^{\dagger}
\partial_k \xi ) d^3 x \nn\\
&= \dfrac{g^2}{32\pi^2} \int \epsilon_{ijk} C_i
(\partial_j C_k - \partial_k C_j) d^3x
=mn.
\label{bkqn}
\eea
This is nothing but the Chern-Simon index of the potential
$C_\mu$. Perhaps more importantly, the knot topology 
can also be described by the Chern-Simon index of 
the electromagnetic potential $A_\mu$.
Indeed we have  
\bea 
&Q =  \dfrac{g^2}{8\pi^2} \int \epsilon_{ijk} A_i F_{jk} d^3x
=mn, 
\label{sckqn}
\eea   
which describes the twisted knot topology of the real
(physical) magnetic flux.
It describes the linking number of two
quantized magnetic fluxes $\Phi_{\hat z}$ and 
$\Phi_{\hat \varphi}$.
Obviously two flux rings linked together can not be separated by
any continuous deformation of the field configuration. This
provides the topological stability of the knot.

\begin{figure}[t]
\includegraphics[scale=0.5]{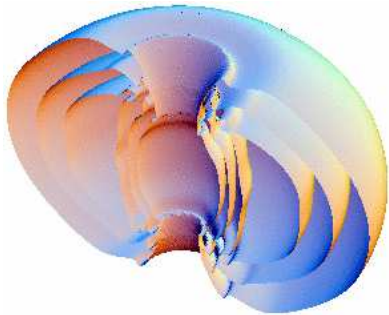}
\caption{(Color online). The energy profile of superconducting
knot with $m=n=1$. Here we have put $g=1,~\lambda=2$, 
$k=0.1~\rho_0$, $~\alpha=0.01$.} 
\label{sck3da}
\end{figure}

Furthermore the topological stability is backed up by 
a dynamical stability. This is because the knot is made of 
two quantized magnetic fluxes
linked together. And the flux trapped inside
of the knot ring can not be squeezed out, and provides
a repulsive force against the collapse of the knot.
This makes the knot dynamically stable.
Another way to understand this dynamical stability
is to notice that the neutral supercurrent
$k_{\hat z}$ of the helical vortex becomes a supercurrent
along the knot, which generates a net
angular momentum around the knot.
And this provides the centrifugal
repulsive force preventing the knot to collapse.
This makes the knot dynamically stable.
We can find the energy profile of
the twisted magnetic vortex ring numerically.
Minimizing the energy we obtain
the energy profile of the knot shown in Fig. \ref{sck3da}.

Clearly the dynamical stability of the knot
comes from the helical structure of the vortex.
But this helical structure is precisely what provides the
knot topology to the vortex ring. The nontrivial topology 
expresses itself
in the form of the helical magnetic flux
which provides the dynamical stability of the knot.
Conversely the helical magnetic flux assures the 
existence of the knot topology which
guarantees the topological stability of the knot.
It is this remarkable
interplay between dynamics and topology which assures
the existence of the stable superconducting knot
in two-gap superconductor.

We close with the following remarks: \\
1. The existence of a knot in two-gap superconductor
has been conjectured before \cite{ijpap,baba}. 
In fact, similar knots have been asserted to exist
almost everywhere in physics, in atomic physics in two-component
Bose-Einstein condensates \cite{ijpap,ruo,pra05},
in nuclear physics in Skyrme theory
\cite{fadd,prl01}, in high energy physics in QCD \cite{plb05}. 
In this paper we have provided a concrete evidence for 
the existence of a stable knot in two-gap superconductor,
which can be interpreted as a twisted magnetic 
vortex ring. \\
2. In this Letter we have concentrated on
the Abelian gauge theory of two-gap superconductor,
which is made of two condensates of same charge.
But we emphasize that a similar knot
should also exist in the non-Abelian gauge theory of
two-gap superconductor which describes two condensates of 
opposite charge \cite{prb05}.
This is because two theories are
mathematically identical to each other. 
This implies that such a knot can also exist in the liquid metallic 
hydrogen (LMH), because it can also be viewed as a non-Abelian 
two-gap superconductor made of opposite charge 
(the proton pairs and the electron pairs) \cite{ash}. \\
3. One should try to construct the topological
knot experimentally in two-gap superconductors.
But before one does that, one must first confirm 
the existence of the non-Abrikosov vortex
which has a non-vanishing condensate (the second component)
at the core. This is very important because this will prove
the existence of a new type of magnetic vortex in 
two-gap superconductor.
With this confirmation one may try
to construct the helical vortex and the knot.
With some experimental
ingenuity one should be able to construct these topological
objects in $\rm MgB_2$ or $\rm LMH$. \\

As we have remarked there is a wide class of realistic potentials
which allows similar non-Abrikosov vortex and topological knot.
The detailed discussion on the subject and its applications
in the presence of a more general potential which includes
the Josephson interaction will be discussed 
separately \cite{cho1}.

{\bf ACKNOWLEDGEMENT}

~~~The work is supported in part by the ABRL Program of
Korea Science and Enginnering Foundation (Grant R14-2003-012-01002-0),
and by the BK21 Project of the Ministry of Education.


\begin{thebibliography}{99}
\bibitem{bec} C. Myatt {\it at al.}, Phys. Rev. Lett. {\bf 78}, 586 (1997);
D. Stamper-Kurn, {\it at al.}, Phys. Rev. Lett. {\bf 80}, 2027 (1998).
\bibitem{sc}J. Nagamatsu et al., Nature {\bf 410}, 63 (2001);
S. L. Bud'ko et al., Phys. Rev. Lett. {\bf 86}, 1877 (2001);
C. U. Jung et al., Appl. Phys. Lett. {\bf 78}, 4157 (2001).
\bibitem{ijpap} Y. M. Cho, cond-mat/012325; 
Int. J. Pure Appl. Phys. {\bf 1}, 246 (2005).
\bibitem{prb05} Y. M. Cho, cond-mat/012492; 
Phys. Rev. {\bf B72}, 212516 (2005).
\bibitem{ruo} U. Al Khawaja and H. Stoof, Nature {\bf 411},
918 (2001); C. Savage and J. Ruostekoski, 
Phys. Rev. Lett. {\bf 91}, 010403 (2003).
\bibitem{pra05} Y. M. Cho, H. J. Khim, and Pengming Zhang, 
Phys. Rev. {\bf A72}, 063603 (2005).
\bibitem{baba} E. Babaev, Phys. Rev. Lett. {\bf 89}, 067001 (2002); 
E. Babaev, L. Faddeev, and A. Niemi, Phys. Rev. {\bf B65}, 100512 (2002).
\bibitem{huang} K. Huang and R. Tipton, Phys. Rev. {\bf D23}, 3050
(1981).
\bibitem{fadd} L. Faddeev and A. Niemi, Nature {\bf 387}, 58 (1997);
R. Battye and P. Sutcliffe, Phys. Rev. Lett. {\bf 81}, 4798 (1998).
\bibitem{prl01} Y. M. Cho, Phys. Rev. Lett. {\bf 87}, 252001 (2001);
Phys. Lett. {\bf B603}, 88 (2004).
\bibitem{plb05} Y. M. Cho, Phys. Lett. {\bf B616}, 101 (2005).
\bibitem{ash} N. Ashcroft, Phys. Rev. Lett. {\bf 92}, 187002 (2004);
E. Babaev, A. Sudbe, and N. Ashcroft, Nature {\bf 431}, 666 (2004).
\bibitem{cho1} Y. M. Cho, and Pengming Zhang, 
to be published.
\end{thebibliography}
\end{document}